\documentclass[preprints,article,accept,pdftex,moreauthors]{Definitions/mdpi}
\firstpage{1}
\articlenumber{zzz-yyy}
\doinum{10.3390/particles1010016 }
\pubvolume{1}
\issuenum{1}
\articlenumber{0}
\pubyear{2026}
\copyrightyear{2026}
\datereceived{01 April 2026}
\pdfoutput=1 

\Title{Revisiting the Rhoades-Ruffini bound}

\Author{David Blaschke $^{1,2,3,*}$\orcidA{} and Adrian Wojcik $^{4}$\orcidB{} }

\address{%
$^{1}$ \quad Institute of Theoretical Physics, University of Wroclaw, Max Born Pl. 9, 50-204 Wroclaw, Poland\\
$^{2}$ \quad Helmholtz-Zentrum Dresden-Rossendorf (HZDR), Bautzner Landstrasse 400, 01328 Dresden, Germany\\
$^{3}$ \quad Center for Advanced Systems Understanding (CASUS), Untermarkt 20, 02826 G\"orlitz, Germany\\
$^{4}$ \quad Institute of Experimental Physics, University of Wroclaw, Max Born Pl. 9, 50-204 Wroclaw, Poland\\
}

\corres{Correspondence: david.blaschke@uwr.edu.pl
}

\abstract{We revisit the derivation of the Rhoades-Ruffini bound on the upper limit for the maximum mass of neutron stars and find that the assumption made there for the onset of an ultimately stiff phase of high-density matter is not stringent. Relaxing this assumption and allowing for an onset of stiff non-nucleonic matter under neutron star constraints at the saturation density or below boost the upper limit of the theoretically possible maximum mass to $4~M_\odot$ or higher, in the mass-gap region between neutron stars and stellar-mass black holes.
We provide a fit formula for the dependence of this upper limit on the speed of sound and the onset density of the deconfinement transition.
}

\keyword{maximum mass limit; constant speed of sound; mass gap}

\protect 
\begin{document}
\maketitle




\section{Introduction}
\label{sec:intro}

In their seminal paper of 1974, Rhoades and Ruffini came to the conclusion that "the maximum mass of a neutron star cannot be larger than $3.2~M_\odot$" \cite{Rhoades:1974fn}.
This finding is a stringent consequence of solving the Tolman-Oppenheimer-Volkoff (TOV) equations of general relativistic hydrodynamic stability of spherical masses with an equation of state (EoS) that assumes a transition from known nuclear matter to the stiffest possible EoS being compatible with the causality constraint for the squared speed of sound
$c_s^2\le c^2=1$ at a density $n_{\rm onset}=1.7~n_0$, where
$n_0=0.15$ fm$^{-3}$ is the nuclear saturation density.
This scheme has been applied ever since when modeling or classifying hybrid neutron stars with a core of stiff high-density matter obeying a constant speed of sound (CSS) EoS.
See, e.g., Ref. \cite{Alford:2013aca}.
The CSS model of high-density quark matter has found a justification in microscopic models of color superconducting
quark matter where it has been found that $c_s^2$ is approximately constant and exceeds the value of $1/3$ that characterizes asymptotically conformal quark matter in perturbative QCD \cite{Zdunik:2012dj,Contrera:2022tqh}.

Together with the observation of a lower mass limit for stellar mass black holes in low-mass X-ray binaries \cite{Bailyn:1997xt} and the expected minimal mass of black holes created in failed supernova explosions \cite{Woosley:2002zz} of about $5~M_\odot$, the Rhoades-Ruffini bound on the upper limit for the mass of neutron stars gave rise to the notion of a mass gap.
Since indeed there were no precise measurements of neutron star masses above $2.5~M_\odot$, the range of masses
$2.5 \le M/M_\odot \le 5.0$   has been defined as a mass gap between the heaviest neutron stars and lightest black holes.

With the advent of precise measurements of the mass of compact objects in binary merger events by the analysis of their gravitational wave signals, there is now evidence against a mass gap because it became populated by objects listed in
\cite{deSa:2022qny}.
Now arises the question about the nature of these mass gap objects like the lighter companion of GW190814 with a  mass of $2.59^{+0.09}_{-0.08}~M_\odot$ \cite{LIGOScientific:2020zkf}, the heavier object in the binary merger GW230529 with a mass of $2.5 - 4.5~M_\odot$ \cite{LIGOScientific:2024elc} and also the remnant of the binary neutron star merger GW170817 with a total mass of $2.74~M_\odot$ \cite{LIGOScientific:2017vwq}.

The Rhoades-Ruffini bound at $3.2~M_\odot$ suggests that these objects could in principle be hybrid neutron stars with an exotic core of stiff matter.
Despite this comfortable maximum mass limit, realistic models of hybrid star EoS yield maximum masses below $\sim 2.5~M_\odot$ \cite{Alvarez-Castillo:2016oln, Ayriyan:2021prr}.
This situation has to be confronted with the finding that within the Rhoades-Ruffini setting, maximum masses exceeding the famous bound have been found, e.g., by Kalogera and Baym \cite{Kalogera:1996ci} when lowering the onset density of the phase transition.
In a systematic analysis of mass-radius (M-R) relations for hybrid neutron stars with high-density matter described by a constant speed of sound equation of state, \cite{Cierniak:2021knt} found that the location of the special point in the M-R plane could extend up to $\sim 4~M_\odot$ when the causality limit $c_s^2=1$ was chosen together with an early onset of the phase transition at $n_{\rm onset}=n_0$, implying a maximum mass well above the Rhoades-Ruffini bound.
Recently, in Ref. \cite{Hippert:2024hum} the authors have considered an upper limit for the squared speed of sound
$c_s^2\le 0.781$ from bounds on hydrodynamic transport coefficients and demonstrated that even under this restriction, the Rhoades-Ruffini bound on the maximum mass can be exceeded so that it is not to be considered as a strict upper limit.

This situation calls for a systematic reinvestigation of the Rhoades-Ruffini bound which we will perform in the present work.

\section{EoS and TOV equation}
\label{sec:eos+tov}

\subsection{Hybrid Equation of State}
\label{ssec:eos}
The generic neutron star EoS should consist of a well-known part that describes the nuclear matter phase from very low densities in the outer crust, where it is clustered in nuclei (A-e phase), over the neutron drip defining the transition to the inner crust (A-e-n phase), to the outer core made of nuclear matter.
For this matter we shall employ the relativistic density functional EoS named "DD2npY-T" (in the following: DD2npY), extended to include the crust matter, see \cite{Shahrbaf:2022upc} for details.
In the following, we denote this EoS for the low-density phase of neutron  star matter as $P_H(\mu)$.

Following Rhoades and Ruffini \cite{Rhoades:1974fn}, for the unknown high-density phase of the inner core matter, however, one assumes the stiffest possible EoS that fulfills the causality constraint that its speed of sound shall not exceed the speed of light ($\hbar=c=1$)
$c_s=\sqrt{dP/d\varepsilon} \le 1$.

For the high-density phase, the class of CSS matter models will be used, 
\begin{equation}
    P(\mu)= A\left(\frac{\mu}{\mu_0}\right)^{1+c^{-2}_s} - B,
\end{equation}
where the model parameters $A$, $B$ and $c_s^2$
are constant and as a scale for the chemical potential we choose the value at the saturation density, $\mu_0 = 971.3$ MeV.
The baryon density $n$ follows from the canonical relation
\begin{equation}
    n(\mu)=\frac{dP}{d\mu}=\left(1+c_s^{-2}\right)\frac{A}{\mu_0}\left(\frac{\mu}{\mu_0}\right)^{c_s^{-2}}.
\end{equation}
Using the above, we arrive at the energy density
\begin{equation}
    \varepsilon=\mu n - P= B+ c_s^{-2}A\left(\frac{\mu}{\mu_0}\right)^{1+c_s^{-2}}.
\end{equation}
The pressure and energy density are related as
\begin{equation}
    P=c_s^2 \varepsilon - \left(1+c_s^2\right)B.
\end{equation}
The pressure slope parameter $A$ does not affect the relation
between pressure and energy density, but its values should be
limited to a range that produces a non-negative density jump at
the phase transition.

The phase transition between the outer hadronic (H) and the inner quark (Q) core matter is obtained by a Maxwell construction
\begin{equation}
    P_H(\mu_c)=P_Q(\mu_c)=P_c,
\end{equation}
where thermodynamic stability requires that the density jump $\Delta n$ at the transition for $\mu=\mu_c$ is nonnegative
\begin{equation}
    \Delta n= \frac{dP_Q}{d\mu}\bigg|_{\mu=\mu_c}-\frac{dP_H}{d\mu}\bigg|_{\mu=\mu_c}\ge 0~.
\end{equation}

Following the setup of Rhoades and Ruffini, we consider the degenerate Maxwell construction case for which $\Delta n=0$, so that $n_H(\mu_c)=n_Q(\mu_c)=n_c$.
In that case, by chosing the values of the hadronic EoS at the transition: $P_c$, $\mu_c$, and $n_c$, one can determine the parameters of the CSS quark matter EoS required to obtain that transition
\begin{eqnarray}
    A&=&\frac{\mu_c n_c}{1+c_s^{-2}}\left(\frac{\mu_0}{\mu_c}\right)^{1+c_s^{-2}}\\
    B&=&\frac{\mu_c n_c}{1+c_s^{-2}}-P_c~.
\end{eqnarray}
In table \ref{tab:my_label} we present the parameters of the CSS quark matter EoS for the case of a phase transition at the saturation density $n_c=n_{\rm onset}=n_0$ and, as in the work of Rhoades and Ruffini, for $n_c=n_{\rm onset}=1.7~n_0$.
\begin{table}[thb]
    \centering
    \begin{tabular}{c||c|c||c|c}
    \hline
   \multirow{3}{*}{$c_s^2$}  & \multicolumn{2}{c||}{$n_{\rm{onset}}=n_0$}&\multicolumn{2}{c}{$n_{\rm{onset}}=1.7~n_0$} \\
   &\multicolumn{2}{c||}{$\mu_c=971.3$ MeV, $P_c=2.56$ MeV/fm$^{3}$}&\multicolumn{2}{c}{$\mu_c=1030.6$ MeV, $P_c=15.13$ MeV/fm$^{3}$}
   \\ \cline{2-5}& A [MeV/fm$^3$]& B [MeV/fm$^3$]& A [MeV/fm$^3$]& B [MeV/fm$^3$]\\
       \hline
        1.0 & 73.50& 70.90& 117.50& 117.15\\
        0.9 & 69.64& 67.08& 110.59& 110.19\\
        0.8 & 65.34& 62.78& 102.91& 102.45\\
        0.7 & 60.53& 57.97& 94.34& 93.81\\
        0.6 & 55.13& 52.57& 84.72& 84.08\\
        0.5 & 49.00& 46.44& 73.83& 73.06\\
        0.4 & 42.00& 39.44& 61.44& 60.46\\
        0.33 &36.48 &33.92 & 51.70& 50.51\\
        0.3 & 33.93& 31.36& 47.23& 45.92\\
\hline
\end{tabular}
    \caption{Parameters $A$ and $B$ of the CSS quark matter EoS for the degenerate Maxwell transition  ($\Delta n=0$) with the hadronic matter EoS DD2npY at $n_{\rm onset}=n_0$ and $n_{\rm onset}=1.7n_0$ for different values of the constant squared speed of sound $c_s^2$.}
    \label{tab:my_label}
\end{table}

\begin{figure}[!h]
    \centering
    \includegraphics[width=\linewidth]{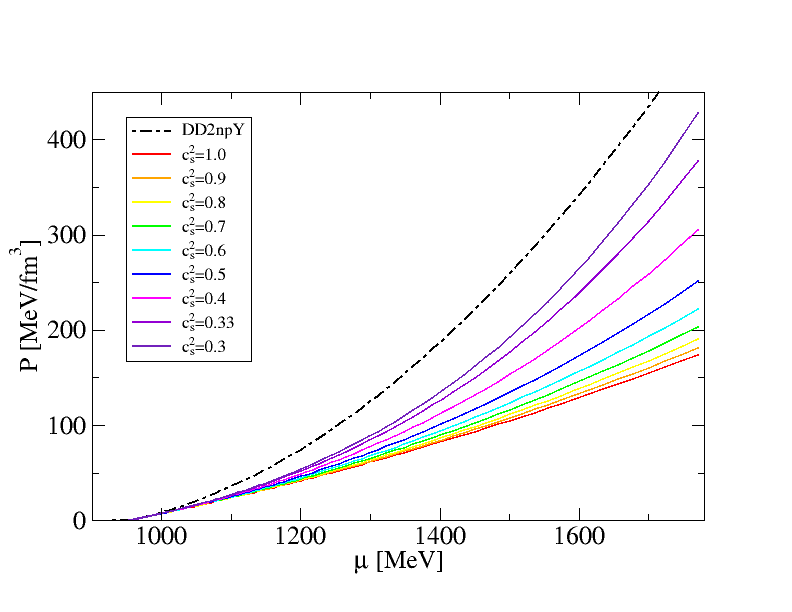}
    \caption{Pressure vs. chemical potential for the hadronic baseline EoS DD2npY (dash-dotted line) and CSS quark matter with different values of the constant sound speed squared (different colored solid lines). The deconfinement onset density $n_{\mathrm{onset}}=n_0=0.15$ fm$^{-3}$ while the density jump at the transition vanishes, $\Delta n=0$. }
    \label{fig:P-mu}
\end{figure}

In Fig. \ref{fig:P-mu} we show the hybrid EoS with with
$n_{\rm onset}=n_0=0.15$ fm$^{-3}$ and $\Delta n=0$ for different values of the squared sound speed in the range
$0.3 \le c_s^2 \le 1.0$.
We note that the lowest curve in this diagram  ($c_s^2 = 1.0$) corresponds to the stiffest EoS which entails the largest maximum mass of the corresponding hybrid star sequence.
This conclusion will be obtained from solving the TOV equations, see the next section.

\subsection{TOV equations}
\label{ssec:tov}
The TOV equations are obtained within general relativity theory under the assumption of a static, spherically symmetric mass distribution in gravitational equilibrium, such as a neutron star,
\begin{equation}
\frac{dP(r)}{dr} = -\frac{G \left[\varepsilon(r) + P(r)/c^2 \right] \left[m(r) + 4\pi r^3 P(r)/c^2 \right]}{r^2 \left[1 - \frac{2Gm(r)}{rc^2} \right]}~,
\end{equation}
where $P(r)$ and $\varepsilon(r)$ are the profiles of pressure and energy density throughout the star and $G$ is the Newtonian gravitational constant.
The mass function $m(r)$ describes the gravitational mass enclosed u to the distance $r$ from the center of the star configuration. It satisfies the equation
\begin{equation}
\frac{dm(r)}{dr} = 4\pi r^2 \varepsilon(r)~.
\end{equation}
The TOV equations ensure that at each layer of the star, pressure supports it against collapse. When the EoS can no longer support the weight (i.e., too high central density), no static solution exists — leading to gravitational collapse (e.g. to a black hole).

\begin{figure}[thb]
    \centering
    \includegraphics[width=0.9\linewidth]{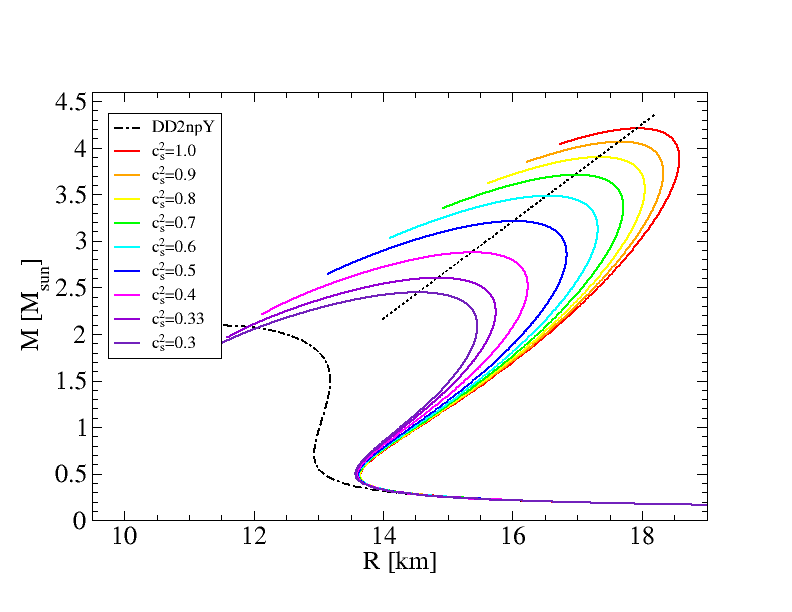}
    \caption{Mass-Radius relations for hybrid stars with onset density $n_{\rm onset}=0.15$ fm$^{-3}$ (solid lines) for different values of the constant squared speed of sound. The maximum mass configurations lie on the dotted line described by the linear fit formula \eqref{eq:Mmax-R}.
    For comparison, the sequence of purely hadronic neutron stars described by the DD2npY-T EoS is shown by a dashed black line. }
    \label{fig:MR_MaxLine}
\end{figure}

\begin{figure}[thb]
    \centering
    \includegraphics[width=0.9\linewidth]{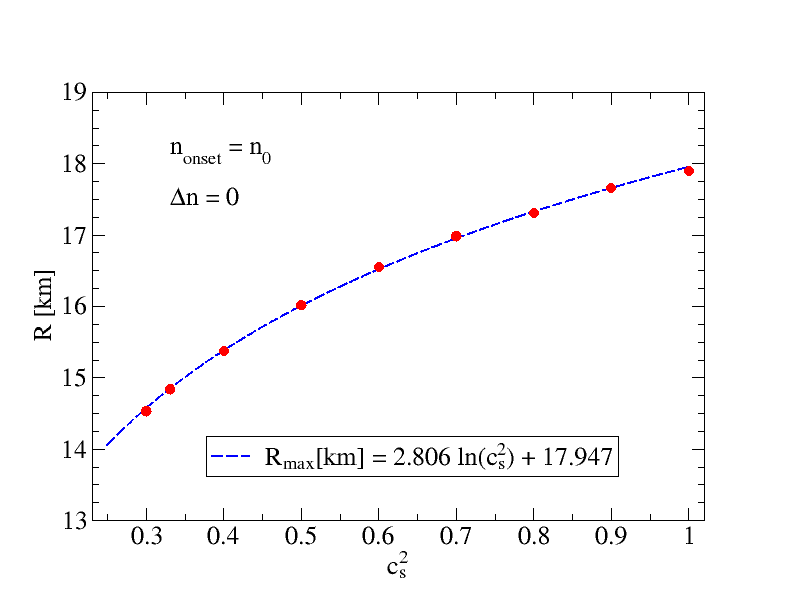}
    \caption{Radius $R=R_{\rm max}$ of the maximum mass configuration as a function of the squared sound speed $c_s^2$, when the onset of deconfinement is at
    $n_{\rm onset}=n_0=0.15$ fm$^{-3}$ and the transition degenerates to a crossover with vanishing density jump $\Delta n=0$.}
    \label{fig:R-c}
\end{figure}

Solutions of the TOV equations with the hybrid EoS described in subsection \ref{ssec:eos} are shown in Fig. \ref{fig:MR_MaxLine} for a fixed onset density of the phase transition at
$n_{\rm onset}=0.15$ fm$^{-3}$ and different examples for the value of the squared speed of sound $c_s^2$.
Increasing the value of $c_s^2$ stiffens the high-density phase of the EoS and leads to an increase in the maximum mass $M_{\rm max}$ and the radius $R_{\rm max}$ where it is attained.
We note that the maximum mass for $c_s^2=1$ grossly exceeds the Rhoades-Ruffini bound of $M_{\rm max}^{\rm RR}=3.2~M_\odot$.
The maximum masses follow a linear dependence on the radius
\begin{equation}
\label{eq:Mmax-R}
    M_{\rm max}[M_\odot]= 0.5217\cdot R_{\rm max}[{\rm km}] - 5.1403~,
\end{equation}
where the location $R_{\rm max}$ of the maximum mass depends on $c_s^2$ as (see Fig. \ref{fig:R-c})
\begin{equation}
    R_{\rm max}[{\rm km}]=2.806 \cdot \ln(c_s^2) + 17.947~.
\end{equation}

The question arises where the striking discrepancy with the result of Rhoades and Ruffini comes from. It is easily identified as the value of the onset density for deconfinement, $n_{\rm onset}$, for which Ref. \cite{Rhoades:1974fn} suggests the value $n_{\rm onset}=1.7~n_0$ while we have chosen
$n_{\rm onset}=n_0$.
Our choice is in accordance with the begin of the region called "terra incognita" in Ref. \cite{Ayriyan:2021prr}, see figures 1 and 2 of that reference.

\begin{figure}[thb]
    \centering
    \includegraphics[width=\textwidth]{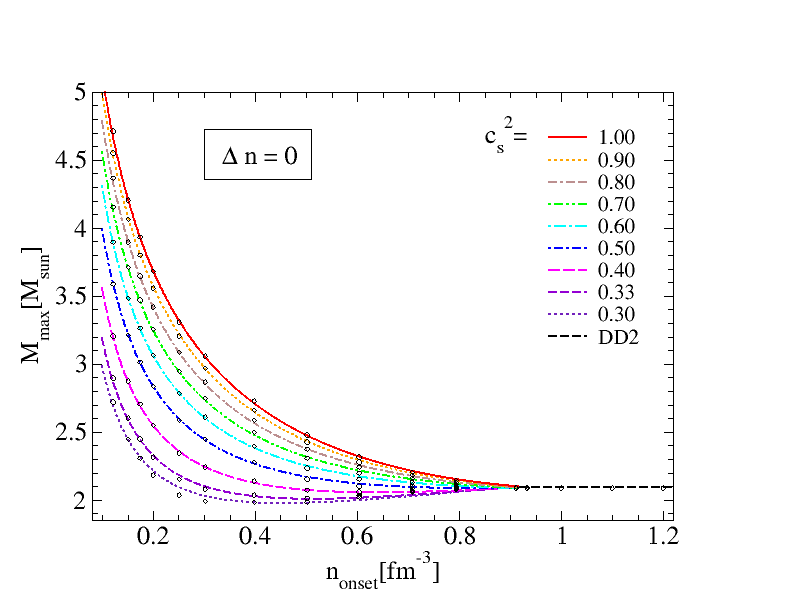}
    \caption{Maximum mass of the hybrid neutron stars as a function of the onset density $n_{\rm onset}$ for selected values
    of the constant squared speed of sound of the high-density phase
    $c_s^2$= 0.30 (violet), 0.33 (magenta), 0.40 (pink), 0.50 (blue), 0.60 (cyan), 0.70 (green), 0.80 (brown), 0.90 (orange) and 1.00 (red). Stable hybrid stars are obtained for onset densities below $n_{\rm onset,max}=0.93$ fm$^{-3}$.}
    \label{fig:Mmax_cs2-onset}
\end{figure}

\section{Results}

In this section, we report our results for the maximum mass upon a variation of both the onset density and the squared sound speed.
The variation of the onset density for $c_s^2=1$ has been performed in the study of Kalogera and Baym \cite{Kalogera:1996ci}, who find for the dependence of the maximum mass on this parameter the behaviour
$M_{\rm max}\propto 1/\sqrt{n_{\rm onset}}$.
The two-dimensional parameter dependence of the Rhoades-Ruffini problem will be reported here for the fist time.

The main result of this paper is that the dependence of the maximum mass of hybrid neutron stars with a stiff high-density (quark matter) phase given by the CSS EoS at the causality limit can be described by the fit formula

\begin{eqnarray}
\label{eq:Mmax}
    M_{\rm max}[M_\odot]&=& M_1(c_s^2)n^{-\alpha(c_s^2)}_{\rm onset}[{\rm fm}^{-3}]
  +M_2(c_s^2)n^{\beta(c_s^2)}_{\rm onset}[{\rm fm}^{-3}]~,
\end{eqnarray}
where $n_{\rm onset}$ is the onset density of the phase transition and $c_s^2$ is the constant squared speed of sound in units of the squared speed of light.

\begin{figure}[thb]
    \centering
    \includegraphics[width=0.9\linewidth]{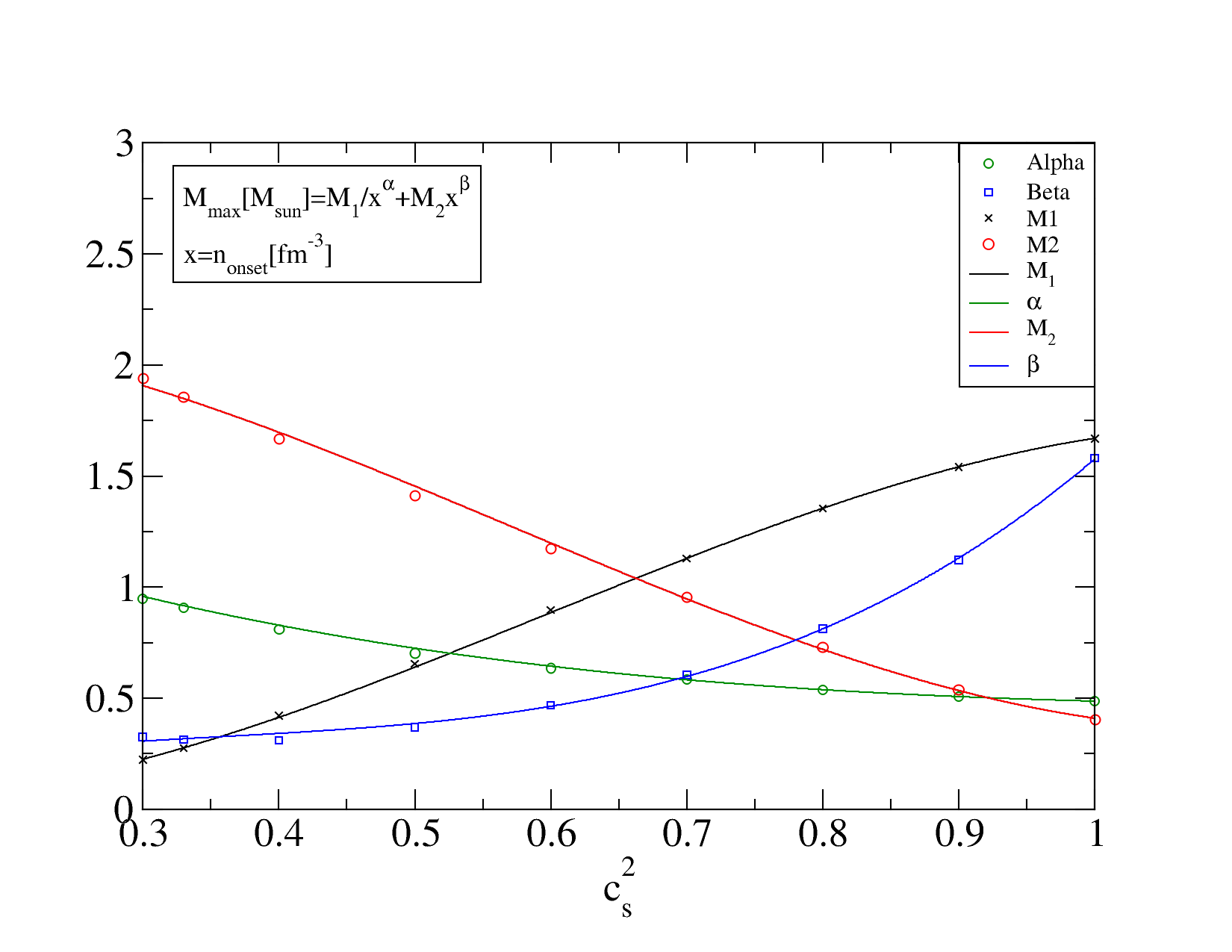}
    \caption{The dependence of the four fit parameters $M_1$, $M_2$, $\alpha$, and $\beta$ of the maximum mass formula \eqref{eq:Mmax} in the text on the squared speed of sound $c_s^2$.}
    \label{fig:parameters}
\end{figure}

In Fig. \ref{fig:Mmax_cs2-onset} we show the maximum mass for different values of the onset density $n_{\rm onset}$ and the squared sound speed  $c_s^2=  0.30~(0.1)~1.0$ and $c_s^2= 0.33$ together with the continuous curves corresponding to the four-parameter fit function \eqref{eq:Mmax} for the dependence of the maximum mass on the onset density.
The coefficients are fitted to third order polynomials of $c_s^2=x$,
\begin{eqnarray}
\label{eq:poly-fit}
    M_1(x)&=& -3.1885 x^3 + 5.7173 x^2 - 0.9362 x + 0.0771,\\
    M_2(x)&=& 3.1017 x^3 -5.3347 x^2 + 0.4842 x + 2.1582,\\
    \alpha(x)&=& -0.5108 x^3 + 1.9049 x^2 - 2.4398 x + 1.5323,\\
    \beta(x)&=& 3.9379 x^3 -4.2532 x^2 + 1.8702 x + 0.0217.
\end{eqnarray}
The fitted values for these four parameters are shown in Fig. \ref{fig:parameters} together with the polynomial fits
\eqref{eq:poly-fit} of their dependence on $c_s^2$.

We observe that for the stiffest EoS at the causality limit $c_s^2=1$, the attainable maximum mass strongly increases when the onset density of the phase transition is lowered.
Therefore, answering the question for a theoretical upper limit on the maximum mass of (hybrid) neutron stars implies answering the question for the minimal possible onset density.
The celebrated Rhoades-Ruffini bound \cite{Rhoades:1974fn} on the upper limit of the maximum mass $M_{\rm max}^{\rm RR}=3.2~M_\odot$ has been obtained by choosing $n_{\rm onset}^{\rm RR} = 0.275$ fm$^{-3}$ without proper justification
\footnote{In Ref.~\cite{Rhoades:1974fn} the mass density at the onset of the high-density CSS phase has been set to $\varrho_{\rm onset}^{\rm RR} = 4.6 \cdot 10^{14}$ g/cm$^3$. For a saturation density $\varrho_{\rm sat} = 2.7 \cdot 10^{14}$ g/cm$^3$, this corresponds to a mass density at the onset of
$\varrho_{\rm onset}^{\rm RR} = 1.7\, \varrho_{\rm sat}$.}.
Since the ratio of onset densities in symmetric nuclear matter and in neutron star matter can be very large, reaching values of 5 to 9 \cite{Panasiuk:2025}, a restriction for the onset density in symmetric matter
$n_{\rm onset}^{\rm SNM} \gtrsim 3\, n_{\rm sat}$,
motivated by the absence of signals of deconfinement in low-energy heavy-ion collisions, would still not exclude that the onset of deconfinement under neutron star conditions may occur at subnuclear densities!
One can read off from Fig. \ref{fig:Mmax_cs2-onset} that in this case the Rhoades-Ruffini bound becomes obsolete and has to be replaced by a new upper limit on the maximum mass that takes into account its dependence on the onset density which can actually lie below the value of the saturation density, so that $M_{\rm max}^{\rm CSS}\gtrsim 4.0~M_\odot$.

The maximum hybrid star masses obtained from Eq. \eqref{eq:Mmax} for different onset densities at the phenomenological upper limit $c_s^2=0.781$ argued for in Ref. \cite{Hippert:2024hum} reproduce well the values found in that paper.

\begin{figure*}[!h]
    \centering
    \includegraphics[width=\linewidth]{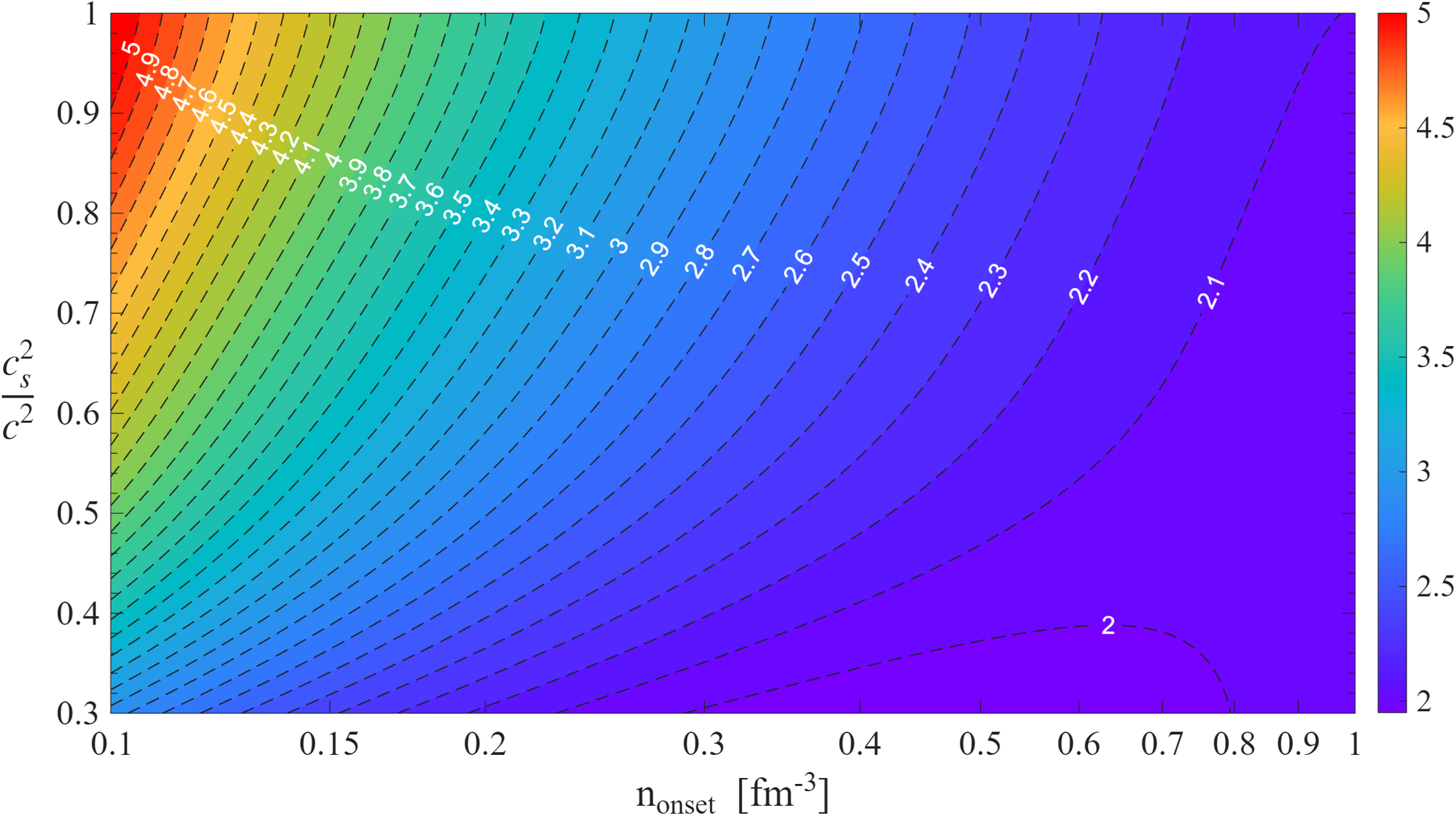}
    \caption{Lines of constant maximum mass of hybrid neutron stars in the plane of squared speed of sound 
    and onset density of the stiff high-density phase.}
    \label{fig:cs2-nonset}
\end{figure*}

In Figure \ref{fig:cs2-nonset} we summarize the main finding of this work that the upper limit of the maximum mass for a hybrid neutron star with a CSS high-density EoS does not only depend on the value of $c_s^2$ that is bounded by causality to $c_s^2\le 1$, but also strongly depends on the lower limit for the onset density of deconfinement which is not well known.

We argue that under neutron star conditions the onset density of deconfinement can be even below the saturation density so that the upper limit on the maximum mass would exceed $4~M_\odot$.
One can read off from Figure \ref{fig:cs2-nonset} that, e.g., for $n_{\rm onset}^{\rm CSS} = 0.11$ fm$^{-3}$ a maximum mass of $M_{\rm max}=4.0~M_\odot$ would be reached already for $c_s^2=0.66$, i.e. well below that phenomenological upper limit from \cite{Hippert:2024hum} and very close to values obtained for color superconducting quark matter  \cite{Contrera:2022tqh}.

\begin{figure}[thb]
    \centering
    \includegraphics[width=\linewidth]{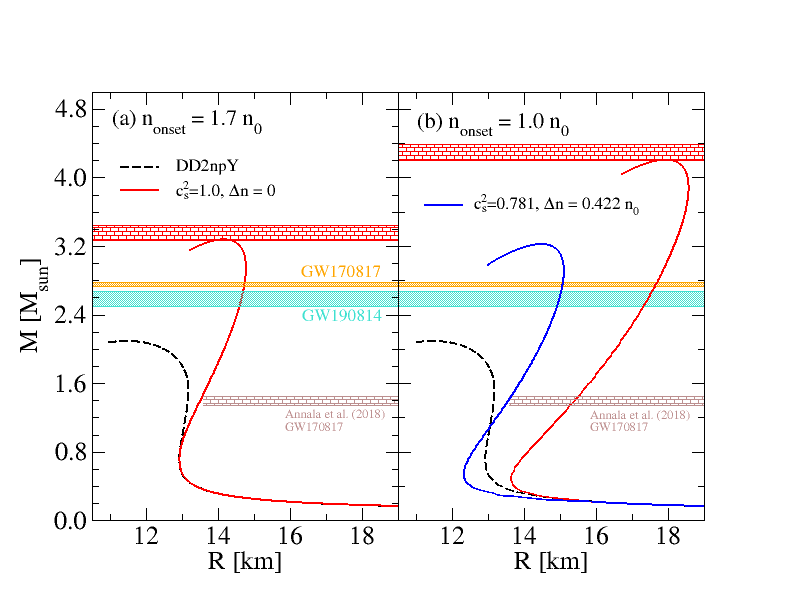}
    \caption{Mass-radius sequences obtained with the hybrid CSS EoS constrained by the upper limit for the radius at $1.4~M_\odot$, $R_{1.4}\le 13.6~M_\odot$
    \cite{Annala:2017llu}.
    }
    \label{fig:M-R-Annala}
\end{figure}


\section{Discussion}

Before concluding, we want to discuss our findings in view of the modern constraint on radii of pulsars from multi-messenger astronomy, see Fig. \ref{fig:M-R-Annala} and return to the question for the nature of mass-gap objects.

In Fig. \ref{fig:M-R-Annala} we give examples for CSS hybrid star EoS parametrizations that would fulfill the modern version \cite{Koehn:2025zzb} of the constraint on the radius at $1.4~M_\odot$ introduced by Annala et al.
\cite{Annala:2017llu}, which amounts to $R_{1.4}\le 13.1~M_\odot$.
We want to underline that even under this constraint maximum masses still reach values similar to the old Rhoades-Ruffini bound in the vicinity of $3~M_\odot$, because of the early onset and a rather stiff high-density phase with $c_s^2\ge 0.65$.
In view of these findings it is conceivable that mass-gap objects in the range $2.5\le M[M_\odot]\le 3.2$ are hybrid neutron stars with color superconducting quark matter cores that would explain the high value of the squared sound speed as well as the early onset of deconfinement.

\section{Conclusions}
\label{sec:concl}

We have repeated and extended the investigation of the theoretical upper limit for the maximum mass of neutron stars in the setting that was defined by Rhoades and Ruffini in 1974.
We relaxed their assumption that the onset density of a maximally stiff form of matter with the speed of sound at the causality limit is at 1.7 times the saturation density. A steep rise of the maximum mass of such hybrid stars is obtained when the onset density is lowered towards the saturation density or even below it.
Treating the onset of deconfinement as a true first-order phase transition with a density jump, one lowers the maximum mass but accommodates the radius constraint from the measured tidal deformability of the binary neutron star merger GW170817 at about 1.4 solar masses.
Ironically, with this setting the new upper limit on the maximum mass of hybrid stars coincides with that suggested by Rhoades and Ruffini, only that with the early onset of a strong phase transition the mass gap objects in the range $2.5\le M[M_\odot]\le 3.2$ could now be explained as hybrid neutron stars with a color superconducting quark matter core.



\vspace{6pt}

\funding{ D.B. was supported by NCN under grant No. 2021/43/P/ST2/03319
}

\dataavailability{The data produced by this research will be shared by the authors upon reasonable request.
}

\acknowledgments{We are grateful to Gordon Baym for pointing out reference \cite{Kalogera:1996ci} to us at the workshop on "The Modern Physics of Compact Stars and Relativistic Gravity" in Yerevan, where these results were first presented in September 2025.
D.B. acknowledges discussions with Oleksii Ivanytskyi and Pavlo Panasiuk about the minimal onset density for the deconfinement phase transition in neutron stars.}

\conflictsofinterest{The authors declare no conflicts of interest.}

\reftitle{References}



\PublishersNote{}
\end{document}